\documentclass[12pt,preprint]{aastex}

\shorttitle{Spectroscopy of Planetary Mass Candidates} 
\shortauthors{Jayawardhana \& Ivanov}

\begin{document}


\title{Spectroscopy of Young Planetary Mass Candidates with Disks}

\author{Ray Jayawardhana} 
\affil{Department of Astronomy \& Astrophysics, University of Toronto,
    Toronto, ON M5S 3H8, Canada}
\email{rayjay@astro.utoronto.ca}

\and 

\author{Valentin D. Ivanov} 
\affil{European Southern Observatory, Ave. Alonso de Cordova 3107, Vitacura, Santiago 19001, Chile} 
\email{vivanov@eso.org}

\begin{abstract}
It is now well established that many young brown dwarfs 
exhibit characteristics similar to classical T Tauri stars, 
including infrared excess from disks and emission lines                                  
related to accretion. Whether the same holds true for even 
lower mass objects, namely those near and below the 
Deuterium-burning limit, is an important question. Here we 
present optical spectra of six isolated planetary mass
candidates in Chamaeleon II, Lupus I and Ophiuchus                                        
star-forming regions, recently identified by Allers and 
collaborators to harbor substantial mid-infrared excesses. 
Our spectra, from ESO's Very Large Telescope and New 
Technology Telescope, show that four of the targets have spectral 
types in the $\sim$M9-L1 range, and three of those also exhibit 
H$\alpha$. Their luminosities are consistent with masses of 
$\sim$5-15 M$_{Jupiter}$ according to models of Chabrier, Baraffe 
and co-workers, thus placing these four objects among the lowest mass 
brown dwarfs known to be surrounded by circum-sub-stellar disks. 
Our findings bolster the idea that free-floating planetary mass 
objects could have infancies remarkably similar to those of 
Sun-like stars and suggest the intriguing possibility of 
planet formation around primaries whose masses are comparable 
to those of extra-solar giant planets. Another target appears 
to be a brown dwarf ($\sim$M8) with prominent H$\alpha$ emission, 
possibly arising from accretion. The sixth candidate is 
likely a background source, underlining the need for 
spectroscopic confirmation. 
\end{abstract}


\section{Introduction}
Thanks to extensive studies conducted in the past five years,
it is now clear that many (if not most) young brown dwarfs
undergo a classical T Tauri phase similar to that of solar mass
stars. Evidence for dusty disks around young sub-stellar objects
come from infrared and millimeter excess (e.g., Natta et al. 2002; 
Jayawardhana et al. 2003; Scholz et al. 2006). Broad and asymmetric 
emission lines, such as H$\alpha$, indicate on-going, and often 
variable, accretion (e.g., Jayawardhana, Mohanty \& Basri 2003; 
Natta et al. 2004; Mohanty et al. 2005; Scholz et al. 2005). 
Forbidden line emission, thought to arise in outflows
or winds, is also seen in some cases (Fern\'andez \& Comer\'on 
2001; Barrado y Navascu\'es \& Jayawardhana 2004; Whelan 
et al. 2005). These striking similarities between young low-mass 
stars and objects near and below the sub-stellar boundary are 
sometimes taken as evidence for a common formation scenario.

Meanwhile, a few dozen free-floating objects with inferred 
masses near and below the Deuterium-burning limit of $\sim$12  
Jupiter masses (Chabrier et al. 2000a) 
have been identified in the $\sigma$ Orionis region 
(Zapatero Osorio et al. 2000) and the Orion 
Nebula Cluster (Lucas and Roche 2000; Lucas, Roche and Tamura 
2005). Low-resolution spectra have confirmed cluster membership 
and ultra low masses for some of them according to evolutionary 
models (Barrado y Navascu\'es et al. 2001; Mart\`in et al. 2001; 
Lucas et al. 2001). These `isolated planetary mass objects' 
(`IPMOs', `planemos') or `sub-brown dwarfs', as they are 
sometimes called, represent the bottom end of the stellar 
initial mass function (IMF). Thus, any 
successful theory of star formation must account for their
origin as well. However, little is known about the 
properties of `planemos' because they are very faint at 
the distances of $\sigma$ Ori and ONC ($\sim$350-450 pc), and 
only a few more have been identified in closer star forming 
regions. In particular, it is extremely interesting to 
investigate whether they also harbor accretion disks, just 
like many of the higher mass young brown dwarfs.  At least 
in a few cases, there is already evidence of mid-infrared 
excess (e.g., Testi et al. 2002; Luhman et al. 2005a, 2005b) 
and strong H$\alpha$ emission from objects near and below 
the Deuterium-burning limit (Barrado y Navascu\'es et al. 
2001, 2002). 

Now, by combining ground-based optical and near-infrared 
photometry with {\it Spitzer} ``Cores to Disks'' Legacy Survey data, 
Allers et al. (2006) have identified six new candidate planemos 
in three nearby regions, at distances $\sim$150 pc. What's 
more, based on their infrared excess, these objects appear to 
be surrounded by dusty disks. However, spectroscopy is 
essential to determine their true nature, in particular whether 
they indeed have late spectral types, and thus relatively cool 
temperatures and very low masses. That is the goal of the 
present {\it Letter}. 

\section{Observations and Data Reduction}
The spectra of targets 1, 5 and 17 were obtained with FORS2 
(FOcal Reducer/low dispersion Spectrograph; Appenzeller et  
al. 1998) at the European Southern Observatory's (ESO) 
Very Large Telescope on Paranal, and for targets 11, 12 
and 18 with EMMI (ESO Multi-Mode Instrument; Dekker, 
Delabre \& D'Odorico 1986) at the ESO New Technology 
Telescope on La Silla. The summary observing log is given 
in Table 1. 

The FORS2 observations used grism GRIS\_150I+27 with order   
sorting filter OG590+32, SR collimator, and 1.3$\arcsec$ wide 
slit, yielding resolution of R$\sim$260 over 600-1100\,nm 
wavelength range. For each target we obtained two spectra of
1347\,sec each.            
  
The EMMI observations were carried out in RILD mode, with 
Grism\,\#7 and order sorting filter OG530\#645, binning 
2$\times$2 and 1.5$\arcsec$ wide slit, yielding resolution of 
R$\sim$280 over 530-1000\,nm wavelength range. We obtained 
one 1268 sec exposure of target 11, one 600 sec exposure of 
target 18, and two 900 sec exposures of target 12. 

The data reduction included the usual steps for long slit 
optical spectrocopy: first bias subtraction and flat fielding. 
Next, we extracted 1-dimensional spectra and removed the sky 
emission lines, interpolating between the sky spectra on both 
sides of the target spectrum with the {\it IRAF} task 
{\it apall} (Tody et al. 1993). After wavelength calibration 
with arc frames, we corrected for the instrument efficiency with 
observations of the spectrophotometric standards EG\,274 and 
LTT\,6248 (Hamuy et al. 1992, 1994), for FORS2 and for EMMI 
observations, respectively.               

\section{Results} 
Fig. 1 shows our spectra of the six candidates, smoothed by 
a 3-pixel boxcar and dereddened using extinction values 
derived by Allers et al. (2006) from multi-color photometry 
and comparison to colors of known young M-type objects. 
(Their reported uncertainty in A$_V$ is $\pm$2 magnitudes.) 
Five of the objects have spectra consistent with being 
late-type stars or brown dwarfs, while the sixth (\#12) is 
most likely a background source, possibly a galaxy or a 
quasar. In fact, the 
$I$-band acquisition image shows some nebulosity surrounding it. 
Based on its very low luminosity, candidate \#12 was inferred 
to have a mass of only 7 M$_{Jupiter}$ if it were a member of 
the Ophiuchus star-forming region (Allers et al. 2006), but our 
spectrum shows that it is unlikely to be a sub-stellar object; 
this result once again underlines that spectroscopy is essential 
to determine the nature of planetary mass candidates. 

For the remaining five candiates, in Fig. 1 we also plot 
spectra of cool field stars from Kirkpatrick et al. (1999) 
for comparison. This allows us to find the closest match 
spectral type for each of the candidates by examining a 
variety of features (e.g., TiO and VO absorption troughs) 
as well as the overall shape of the spectrum. In practice, 
we overlaid (scaled) comparison spectra for a number of 
spectral sub-types along with each target's spectrum  
and chose the ``best-match''; We repeated this procedure                          
with spectra of several standards of each sub-type for each 
target. However, we note the well known difficulties of 
deriving precise and reliable spectral types for the coolest 
young objects, especially since the classification is based 
on objects with very different surface gravities (e.g., 
Testi et al. 2002). With these caveats in mind, we report 
our best-fit spectral types for these five candidates in 
Table 2. For \#1, \#11 and \#18, we expect our derived 
classification to be accurate to within about one sub-type, 
while for \#5 and \#17 the uncertainty may be closer to 
two sub-types. 

In Fig. 2, we show a zoom of the spectral region around 
H$\alpha$ for the five late type objects. Three of them 
(\#01, \#11, \#18) show clear emission. A fourth object 
(\#05) does have H$\alpha$ as well, though the detection 
is less reliable due to low signal-to-noise. The sixth 
object, \#17, does not show emission in our (noisy) 
spectrum. Given the poor detection of the 
continuum in our spectra, it is difficult to measure the 
H$\alpha$ equivalent widths (EWs) reliably, so we report 
minimum EWs in Table 2. 

\section{Discussion}
We find that five of the six candidates have spectral types 
in the late M to early L range, as would be expected for low 
mass brown dwarfs and planetary mass objects, while a sixth 
(\#12) is probably a background source. Given that the spectral 
types we derive and the effective temperatures found by Allers 
et al. from photometry are consistent within the errors, these 
five objects are likely to be at the distances of the 
Chamaeleon II, Lupus I and Ophiuchus clouds. Their 
{\it Spitzer}-detected mid-infrared excesses, modeled as emission 
from dusty disks by Allers et al., provide strong evidence 
of youth. Here it is interesting to note that the only object 
in their sample for which Allers and co-workers were not able 
to fit the spectral energy distribution (SED) well with a flat or 
flared disk model is \#12, which we now find to be a likely 
background source. The large minimum H$\alpha$ EWs we find in 
at least three of the five late-type objects bolsters the case 
for their youth. H$\alpha$ could originate from disk accretion or 
chromospheric activity in young low-mass objects (e.g., 
Jayawardhana, Mohanty \& Basri 2002), and it is often difficult 
to distinguish definitively between these two possibilities with 
low-resolution spectra. However, according to the criterion 
developed by Barrado y Navascu\'es \& Mart\`in (2003), 
three of these objects, if not four, could well be accreting. 
The lack of clear H$\alpha$ emission in \#17 does not necessarily 
rule out youth; young L-type objects may not show H$\alpha$ unless 
they are accreting (Barrado y Navascu\'es et al. 2001, 2002). 

There are many uncertainties involved in deriving the masses of 
young very low mass objects, given the difficulties of determining 
their ages and distances reliably as well as the uncertainties in the 
spectral type to T$_{eff}$ conversion and in the evolutionary models 
themselves (e.g., Baraffe et al. 2002; Mohanty, Jayawardhana \& Basri 2004). 
Allers et al. (2006) estimate masses for these candidates by matching 
their source luminosities to the widely used isochrones of Baraffe et 
al. (2001; 2003), and assuming ages of 1 Myr for Ophiuchus and Lupus I and 
3 Myr for Chamaeleon II. They also point out that three of the objects 
we have now confirmed as late type (\#01, \#05, \#17) have luminosities 
equivalent (within the errors) to the lowest luminosity young brown 
dwarf with mid-infrared excess reported previously (Luhman et al. 2005b).  

While we do not think an exhaustive analysis is warranted, given 
the uncertainties involved, we can at least check whether the spectral 
types we have derived are roughly consistent with the temperatures derived 
by Allers et al. from source SEDs. Thus, in Table 2 we report 
effective temperatures for our targets, based on the spectral type to 
T$_{eff}$ conversion scale of Mart\`in et al. (1999), along with those 
from Allers et al. The comparison shows that the 
two sets of derived T$_{eff}$ indeed agree within $\pm$100 K for four 
out of five objects. For the fifth, \#18, we find a somewhat earlier 
spectral type, and a correspondingly higher T$_{eff}$, than implied 
by the Allers et al. analysis. 

Fig. 3 shows a Hertzsprung-Russell diagram of the five late-type 
objects. The absolute $K$ magnitude was derived from the apparent 
$K$ magnitude, the distance modulus and the visual absorption ($A_v$) 
given in Allers et al., and adopting the reddening law from Rieke 
\& Lebofsky (1985). The effective temperatures ($T_{eff}$) are 
those reported in Table 2. The isochrones for 1 Myr and 5 Myr 
DUSTY00 models (Chabrier et al. 2000b; Baraffe et al. 2001) 
are also plotted. This figure shows that four of the candidates 
have masses near or below the deuterium-burning limit, while \#18
is a somewhat higher mass brown dwarf. Given the slope of the 
isochrones around the locus occupied by 
these objects and the $\approx \pm$100 K uncertainty in $T_{eff}$, 
this figure suggests an uncertainty in the mass estimates on the 
order of a few Jupiters for four objects and $^{+7}_{-5}$ 
for the fifth, \#18. The ages of all five objects are consistent, 
within the uncertainties, to those assumed by Allers et al. 

The targets \#1, \#5, \#11 and \#17 are among the lowest mass 
objects with dusty disks known to date, and at least three of the 
four also show possible signs of accretion from those disks.  
Whatever their exact masses are, these objects represent the bottom  
end of the stellar IMF. It is worth noting that Mohanty et al. 
(2006) have recently found evidence for an edge-on disk surrounding 
the planetary mass companion to the nearby young brown dwarf 
2MASSW J1207334-393254, which itself is known to harbor a disk. 

Our findings, combined with previous work, 
suggest that some planetary mass objects have characteristics usually 
seen in T Tauri stars and higher mass brown dwarfs, implying strikingly 
similar infancies for our Sun and objects that are some hundred times 
less massive. Thus, a successful theory for star formation should be 
able to account for these similarities in young objects with a wide 
range of masses.  

The shape of the IMF at these lowest masses is not yet well defined 
observationally. Deep, wide-field optical and infrared surveys with 
8-meter class telescopes are needed to investigate this regime 
with larger samples (e.g., Lucas, Roche \& Tamura 2005). With 
current observing facilities, spectroscopic confirmation of very 
low mass candidates is challenging, but not impossible as we have 
demostrated here (also see Lucas et al. 2001; Barrado y Navascu\'es 
et al. 2001). As Allers et al. have shown, {\it Spitzer} could play a pivotal 
role in determining the disk frequency in the planetary mass regime. 

\acknowledgments
This paper is based on observations collected at the European 
Southern Observatory, Chile, under program number 276.C-5050. 
We thank the ESO staff for carrying out the observations in service 
mode, Aleks Scholz for useful discussions, and an anonymous referee 
for suggestions towards improving the manuscript. This research was 
supported by an NSERC grant and University of Toronto startup funds 
to RJ.

\begin{figure}[t]
\begin{center}
\includegraphics[width=15cm,angle=0]{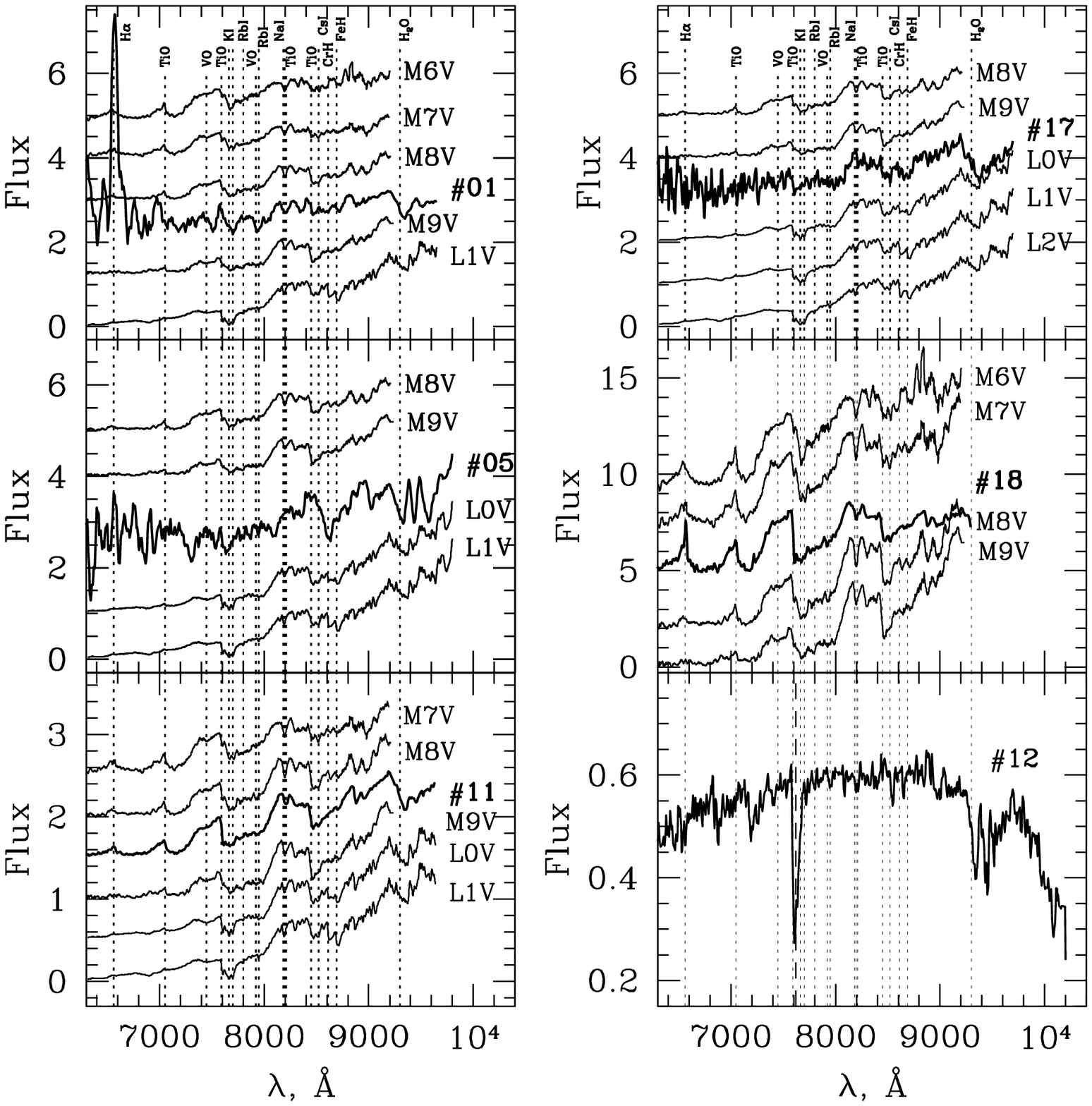}
\caption{Optical spectra of the six `planemo' candidates, along with those 
of comparison objects. Dotted lines mark the location of H$\alpha$ 
and a number of other features while the dashed line in the lower-right 
panel corresponds to a telluric absorption feature.  
\label{fig1}} 
\end{center}
\end{figure}

\clearpage
\newpage

\begin{figure}[t]
\begin{center}
\includegraphics[width=12.5cm,angle=0]{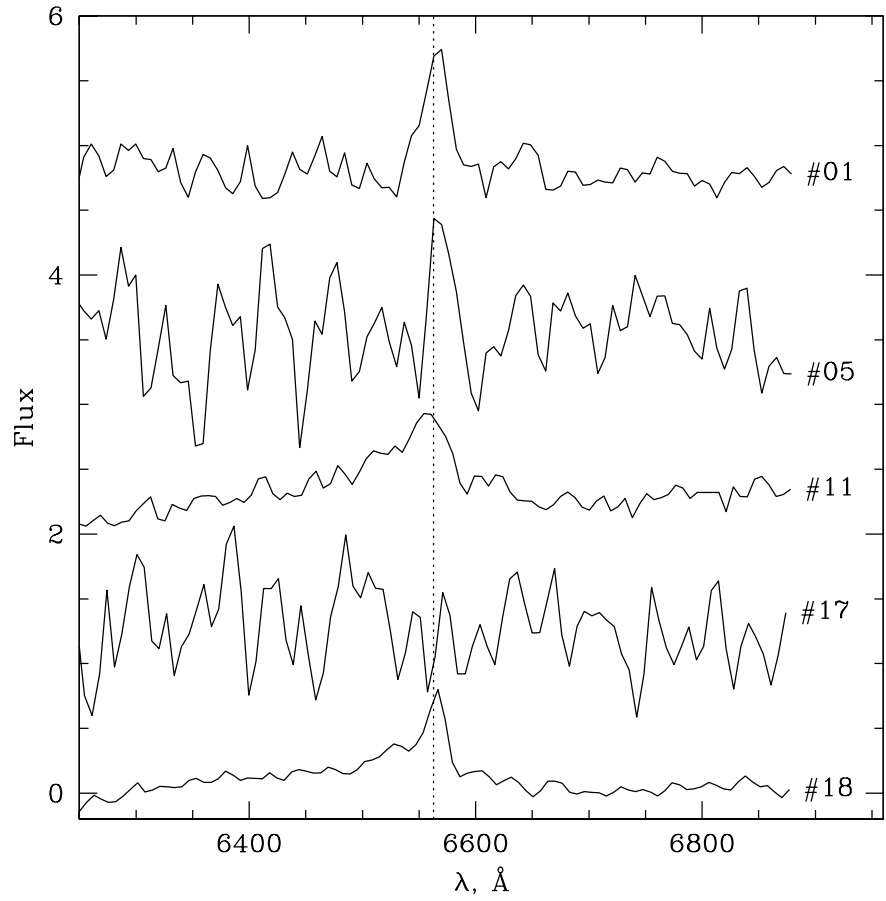}
\caption{Zooming in on the region that includes the H$\alpha$ line for the five late-type objects.  
\label{fig2}} 
\end{center}
\end{figure}

\begin{figure}[t]
\begin{center}
\includegraphics[width=14cm,angle=0]{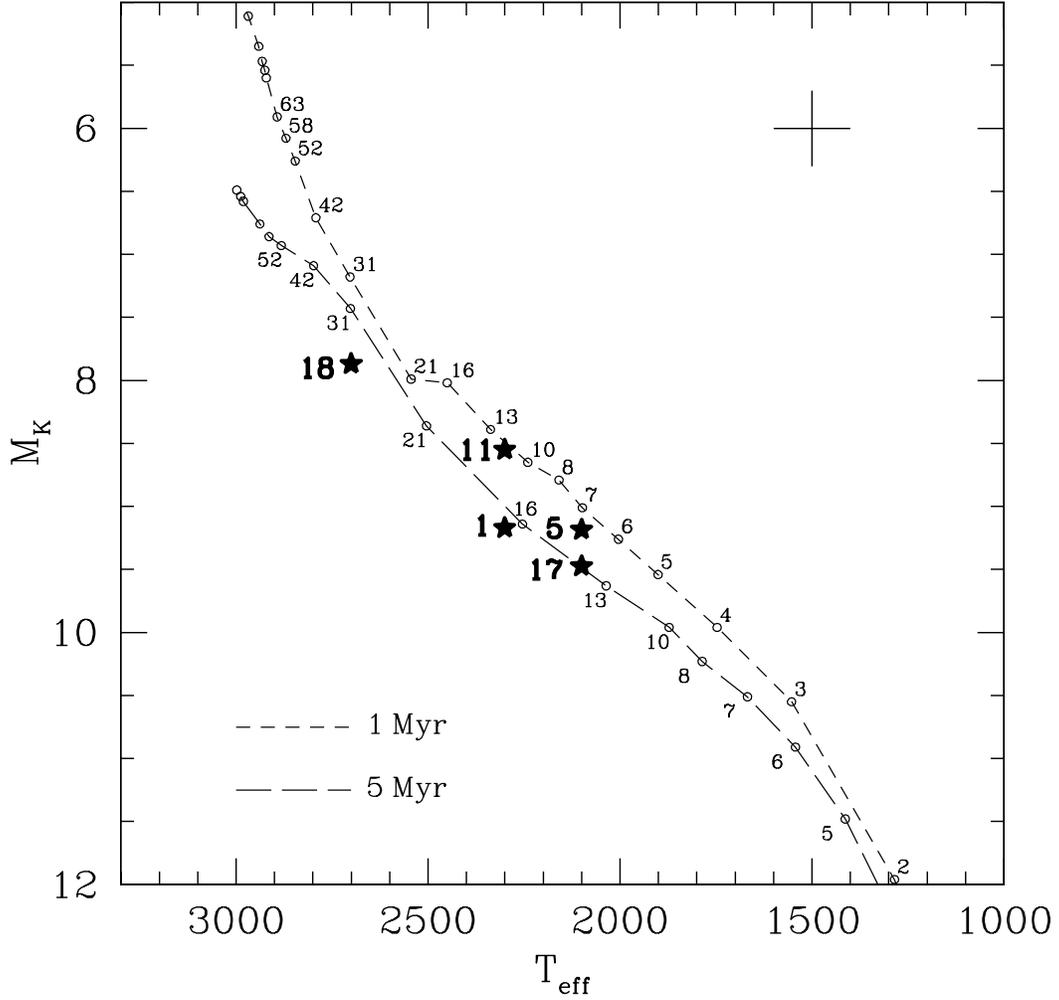}
\caption{Hertzsprung-Russell diagram of the five late-type objects in 
our sample, labelled according to Allers et al. (2006) notation. 
Isochrones for 1 Myr (short-dashed line) and 5 Myr (long-dashed line) 
are plotted, with masses marked in units of Jupiter mass, for  
DUSTY00 models (Chabrier et al. 2000b; Baraffe et al. 2001). 
The typical errorbars are shown in the top right corner. The $M_K$ 
uncertainty combines the photometric errors given in Allers et al. 
with estimates of the errors in the extinctions (0.2 mag in $K$) 
and the distances (15\%), added in quadrature. The typical error in 
$T_{eff}$ is estimated to be $\pm$100 K. 
\label{fig3}}
\end{center}
\end{figure}

\clearpage
\newpage
\begin{deluxetable}{lclc}
\tabletypesize{\scriptsize}
\tablecaption{Summary Observing Log}
\tablewidth{0pt}
\tablehead{
\colhead{Target Name} & \colhead{Target ID\tablenotemark{a}} & \colhead{Date} & {Start of Observations (UT)}} 
\tablecolumns{4}
\startdata 
Cha 125758-770119 & 01 & 2006-03-15 & 06:40:48.797 \\
Cha 130540-773958 & 05 & 2006-03-15 & 07:41:12.119 \\
Oph 162225-240515 & 11 & 2006-03-13 & 06:23:08.854 \\
Oph 162230-232224 & 12 & 2006-03-13 & 07:03:17.530 \\
Lup 153927-344844 & 17 & 2006-03-15 & 08:41:38.721 \\
Lup 154140-334518 & 18 & 2006-03-13 & 06:00:25.611 \\
\enddata 
\tablenotetext{a}{Target ID number is from Allers et al. (2006)}
\end{deluxetable}

\begin{deluxetable}{ccccc}
\tabletypesize{\scriptsize}
\tablecaption{Derived Parameters}
\tablewidth{0pt}
\tablehead{
\colhead{Target} & \colhead{Min.} & \colhead{Spectral} & \colhead{T$_{eff}\pm100$ K} & \colhead{T$_{eff}$ K from}\\
\colhead{ID} & \colhead{EW(H$\alpha$)} & \colhead{Type} & \colhead{from SpT} & \colhead{Allers et al. (2006)}} 
\tablecolumns{5}
\startdata
01 & -90\AA & M9 & 2300  & 2207 \\
05 & -30\AA\tablenotemark{a} & L0 & 2100  & 2100 \\
11 & -50\AA & M9 & 2300  & 2207 \\
17 & -- & L0 & 2100  & 2004 \\
18 & -55\AA & M8 & 2700  & 2400 \\
\enddata
\tablenotetext{a}{Due to poor signal-to-noise in this spectrum, the uncertainty in the EW(H$\alpha$) could be 
as large as $\pm$15\AA.}
\end{deluxetable}


\begin{thebibliography}{}
\bibitem[Allers et al. (2006)]{a06}
Allers, K.N., Kessler-Silacci, J.E., Cieza, L.A., \& Jaffe, D.T.
2006, ApJ, in press

\bibitem[Appenzeller et al. (1998)]{aff98}
Appenzeller, I., et al. 1998, The Messenger 94, 1 

\bibitem[Barrado y Navascu\'es et al. (2001)]{ByN01}
Barrado y Navascu\'es, D., et al. 2001, A\&A, 377, L9

\bibitem[Barrado y Navascu\'es et al. (2002)]{byn02}
Barrado y Navascu\'es, D., et al. 2002, A\&A, 395, 813

\bibitem[Barrado y Navascu\'es \& Jayawardhana (2004)]{ByNJ04}
Barrado y Navascu\'es, D. \& Jayawardhana, R. 2004, \apj, 615, 840 

\bibitem[Barrado y Navascu\'es \& Mart\`in(2003)]{bm03}
Barrado y Navascu\'es, D. \& Mart\`in, E. L. 2003, \aj, 126, 2997

\bibitem[Baraffe et al.(2001)]{b01}
Baraffe, I., et al. 2001, A\&A, 382, 563

\bibitem[Baraffe et al.(2002)]{b02}
Baraffe, I., et al. 2002, A\&A, 382, 563

\bibitem[Baraffe et al.(2003)]{b03}
Baraffe, I., et al. 2003, A\&A, 402, 701

\bibitem[Chabrier et al. (2000a)]{cha00a}
Chabrier, G., et al. 2000a, ApJ, 542, L119 

\bibitem[Chabrier et al. (2000b)]{cha00b}
Chabrier, G., et al. 2000b, ApJ, 542, 464

\bibitem[Dekker et al. (1986)]{ddd86}
Dekker, H., Delabre, B. \& D'Odorico, S. 1986, SPIE 627, 39

\bibitem[Fern\'andez \& Comer\'on (2001)]{fc01}
Fern\'andez, M. \& Comer\'on, F. 2001, A\&A, 380, 264 

\bibitem[Hamuy et al. (1992)]{hamuy92}
Hamuy, M., et al. 1992, PASP, 104, 533 

\bibitem[Hamuy et al. (1994)]{hamuy94}
Hamuy, M., et al. 1994, PASP, 106, 566      

\bibitem[Jayawardhana, Mohanty, \& Basri(2002)]{jmb02}
Jayawardhana, R., Mohanty, S. \& Basri, G. 2002, \apj, 578, L141

\bibitem[Jayawardhana, Mohanty, \& Basri(2003)]{jmb03}
Jayawardhana, R., Mohanty, S. \& Basri, G. 2003, \apj, 592, 282

\bibitem[Jayawardhana et al.(2003)]{jas03}
Jayawardhana, R., et al. 2003, \aj, 126, 1515

\bibitem[Kirkpatrick et al. (1999)]{k99}
Kirkpatrick, J.D., et al. 1999, \apj, 519, 802

\bibitem[Lucas \& Roche (2000)]{lr00}
Lucas, P.W. \& Roche, P.F. 2000, \mnras, 314, 858 

\bibitem[Lucas et al. (2001)]{l01}
Lucas, P.W., et al. 2001, \mnras, 326, 695 

\bibitem[Lucas, Roche \& Tamura (2005)]{lrt05}
Lucas, P.W., Roche, P.F. \& Tamura, M. 2005, \mnras, 361, 211

\bibitem[Luhman et al. (2005a)]{l05a}
Luhman, K.L., et al. 2005a, \apj, 620, L51 

\bibitem[Luhman et al. (2005b)]{l05b}
Luhman, K.L., et al. 2005b, \apj, 635, L93 

\bibitem[Mart\`in et al. (1999)]{m99}
Mart\`in, E.L., et al. 1999, \aj, 118, 2466

\bibitem[Mart\`in et al. (2001)]{m01}
Mart\`in, E.L., et al. 2001, \apj, 558, L117



\bibitem[Mohanty, Jayawardhana, \& Basri(2004)]{mjb04}
Mohanty, S., Jayawardhana, R. \& Basri, G. 2004, \apj, 609, 885 

\bibitem[Mohanty, Jayawardhana, \& Basri(2005)]{mjb05}
Mohanty, S., Jayawardhana, R. \& Basri, G. 2005, \apj, 626, 498

\bibitem[Mohanty et al. (2006)]{m06}
Mohanty, S., Jayawardhana, R., Hu\'elamo, N. \& Mamajek, E.E. 2006, \apj, submitted 

\bibitem[Natta et al.(2002)]{n02}
Natta, A., et al. 2002, A\&A, 393, 597 

\bibitem[Natta et al.(2004)]{ntm04}
Natta, A., et al. 2004, \aap, 424, 603

\bibitem[Rieke \& Lebofsky (1985)]{rl85}
Rieke, G.H. \& Lebofsky, M.J. 1985, ApJ, 288, 618

\bibitem[Scholz, Jayawardhana, \& Brandeker(2005)]{sjb05}
Scholz, A., Jayawardhana, R. \& Brandeker, A. 2005, \apj, 629, L41

\bibitem[Scholz, Jayawardhana, \& Wood (2006)]{sjw06}
Scholz, A., Jayawardhana, R. \& Wood, K. 2006, ApJ, 638, 1056 

\bibitem[Testi et al. (2002)]{t02}
Testi, L., et al. 2002, \apj, 571, L155

\bibitem[Tody et al. (1993)]{tody93}
Tody, D., et al. 1993, in {\it Astronomical Data Analysis Software and Systems II}, 
A.S.P. Conference Ser., Vol 52, eds. R.J. Hanisch, R.J.V. Brissenden, \& J. Barnes, 173 

\bibitem[Whelan et al. (2005)]{whelan05}
Whelan, E.T., et al. 2005, Nature, 435, 652 

\bibitem[Zapatero Osorio et al. (2000)]{z00}
Zapatero Osorio, M.R., et al. 2000, Science, 290, 103 

\end{thebibliography}
\end{document}